# Soft-Matter-Based Topological Vertical Cavity Surface Emitting Lasers

Yu Wang[1,*], Shiqi Xia[1,*], Jingbin Shao[1], Qun Xie[1], Donghao Yang[1], Xinzheng Zhang[1, 2, 4, ‡], Irena Drevensek-Olenik[3, 4], Qiang Wu[1], Zhigang Chen[1, 2, ‡] and Jingjun Xu[1, ‡]

*1* The MOE Key Laboratory of Weak-Light Nonlinear Photonics, TEDA Institute of Applied Physics and School of Physics, Nankai University, Tianjin, 300457, China

*2* Collaborative Innovation Center of Extreme Optics, Shanxi University, Taiyuan, Shanxi 030006, China

*3* Faculty of Mathematics and Physics, University of Ljubljana, and Department of Complex Matter, J. Stefan Institute, SI-1000 Ljubljana, Slovenia

*4* International Sino-Slovenian Joint Research Center on Liquid Crystal Photonics, Nankai University, Tianjin, 300071, China

\* Y. Wang, and S. Xia are co-first authors of the article.

‡ Corresponding author: zxz@nankai.edu.cn; zgchen@nankai.edu.cn; jjxu@nankai.edu.cn

**Abstract:** Polarized topological vertical cavity surface-emitting lasers (VCSELs), as stable and efficient on-chip light sources, play an important role in the next generation of optical storage and optical communications. However, most current topological lasers demand complex design and expensive fabrication processes, and their semiconductor-based structures pose challenges for flexible device applications. By use of an analogy with two-dimensional Semenov insulators in synthetic parametric space, we design and realize a one-dimensional optical superlattice (stacked polymerized cholesteric liquid crystal films and Mylar films), thereby we demonstrate a flexible, low threshold, circularly polarized topological VCSEL with high slope efficiency. We show that such a laser maintains a good single-mode property under low pump power and inherits the transverse spatial profile of the pump laser. Thanks to the soft-matter-based flexibility, our topological VCSEL can be "attached" to substrates of various shapes, enabling desired laser properties and robust beam steering even after undergoing hundreds of bends. Our results may find applications in consumer electronics, laser scanning and displays, as well as wearable devices.

**Introduction**

In recent years, the field of topological physics has advanced rapidly, inspiring researches in many different fields including solid-state physics, cold atoms, photonics, acoustics and mechanics [1, 2, 3, 4, 5, 6]. After the first proposal [7] and pioneering experimental demonstrations [8, 9, 10] in the field of photonics, a variety of topological phenomena have been explored in various systems, including plasmonics [11, 12, 13, 14], metamaterials [15, 16, 17, 18], coupled resonators [19, 20, 21, 22], waveguide arrays [23, 24, 25, 26, 27], polaritonic microcavities [28, 29, 30, 31] and optical systems with synthetic dimensions [32, 33, 34, 35]. New approaches have been developed for implementing nanophotonic devices based on robust topological states [36, 37, 38, 39, 40]. A typical example is the topological insulator lasers – one of the most promising and practical applications of topological photonics – which have been demonstrated in a series of experiments over the past several years [41, 42, 43, 44, 45, 46, 47, 48, 49, 50, 51].

Driven by the practical applications of integrated photonics, there is a growing need for reliable miniatured on-chip laser sources [52, 53]. In particular, vertical cavity surface-emitting lasers (VCSELs) with polarization characteristics have gradually become the core devices for the next generation of optical data storage and optical communications [54, 55, 56], with also a wide range of applications in medical imaging, environmental monitoring, laser scanning, and so on. Recently, topological VCSELs have been proposed and demonstrated, either as a single laser or an arrayed light source, with optimal lasing performance such as in directionality, low-threshold and high-robustness [47, 48, 49, 57, 58]. However, topological VCSELs demonstrated so far are mostly based on semiconductor gain media and solid-state substrates, which often demands complex fabrication techniques, although a perovskite quantum-dot-based topological laser in the visible region has been realized with a lithography-free approach [58]. Due to the widespread demand for laser beam steering technologies and portable laser display devices, it is natural to ask whether a topological VCSEL can be made shape-flexible, lightweight and chip-compatible at low cost.

In the meantime, soft-matter multifunctional photonic materials with self-supporting properties have promoted the development of various wearable devices [59, 60] and have shown superior performance in the field of light field regulation [61]. Cholesteric liquid crystal (CLC), as a typical soft substance, not only has the birefringence of crystals and one dimensional (1D) selectively reflected photonic band gaps, but also has an ordered and controllable molecular arrangement [62]. Taking advantages of such properties, topological lasing has recently been realized by using a 1D polymer-

CLC superlattice, exhibiting properties such as low threshold, tunability, and circular polarization [63]. However, due to the fluidity of the CLC, an in-plane topological laser with hard substrates cannot sustain high pump intensity. To overcome the fluidity and instability while still retain the excellent optical properties of CLCs, polymerized CLC (PCLC) films have attracted increasing attention. Since PCLCs are easy to be processed into thin films, they are ideal for flexible VCSELs, consisting of PCLC Fabry-Perot resonators and organic gain materials. With appropriate implementation, such lasers can be used as thin-film lasers alone or even attached to substrates of any sizes and shapes [64], greatly reducing the device size and increasing the application potential of topological lasers.

In this work, we demonstrate for the first time to our knowledge a circularly polarized flexible topological VCSEL with high slope efficiency. The topological VCSEL is designed based on a 1D optical superlattice with broken inversion symmetry by introducing modulated on-site potentials, which can be analogy with the Semenov insulator in a 2D synthetic parametric space. In particular, our topological VCSEL is prepared by stacking PCLC films and commercial Mylar films, which does not require complex processes such as lithography, deposition and etching. The lasing threshold is as low as about 0.47 μJ ($1.5 \times 10^6$ W/cm$^2$), and the total lasing slope efficiency is 4.18%, much higher than that of the similar PCLC lasers. Such a soft-matter-based topological VCSEL may be an ideal candidate for applications in areas as diverse as artificial intelligence, sensing, virtual and augmented reality displays, and biomedical care devices.

**Results**

**Construction of 1D flexible topological VCSEL based on the Semenov insulator**

Due to the difficulties in breaking the time-reversal and space-inversion symmetries, most topological photonic systems have so far maintained both symmetries. For example, topological photonic structures based on the classical 1D Su-Schrieffer-Heeger (SSH) model satisfy the inversion symmetry and are protected by the chiral symmetry. However, breaking symmetries often causes the optical modes to have non-trivial spatial distributions. For example, by applying a magnetic field to break the time-reversal symmetry, the quantum-like Hall effect in optical systems with strong magneto-optical response could be realized [7]. With the valley degree of freedom of photons, the quantum valley Hall effect can also be realized by using valley photonic crystals breaking the inversion symmetry [65]. Generally, the combination of the time-reversal and space-inversion symmetries protects the gapless properties of Dirac fermions in graphene. However, by introducing atoms with different masses in the hexagonal lattice and breaking the inversion symmetry, an electron energy gap will be created while there are no topological edge states within the bandgap. This system was first discussed by Semenov and is thus called a Semenov insulator [66, 67]. Obviously, Semenov insulators, such as BN, SiC lattices, are insulated both on the whole and at the edges, showing the topologically trivial properties. It is worth noting that splicing BN and NB lattices with different valley Chern numbers will cause a topologically protected mode, which is similar to the quantum valley Hall effect [68, 69].

According to the Hamiltonian described by Dirac equation, it can be found that the on-site potential is equivalent to the mass term of the atom within a unit cell [70]. Recently, Huang et al. designed an acoustic structure based on the discrete spring mass model and modulated the height of the resonator to simulate the disturbed mass [71]. The calculated band structure shows that one interface state could also be generated by splicing two topologically trivial structures with inversion symmetry breaking, and the sound wave was strongly localized at the interface. The similarities between this 1D system and the 2D BN hexagonal lattice enlighten us, whether the 1D equally-coupled and site-potential modulated diatomic chains have similar topological properties with the 2D quantum valley Hall effect.

In general, the dimension of a physical system cannot be larger than its geometric dimensionality. Constructing a synthetic space by coupling states in an artificial lattice or introducing structural parameters can provide an ingenious way to realize a high-dimensional physical system [72, 73, 74]. Here, we constructed a 1D flexible optical superlattice with PCLC films and polymer Mylar films. The synthetic parametric space is formed by introducing the on-site potential modulation, namely adjusting the thickness of the Mylar films, which is similar to Semenov insulators.

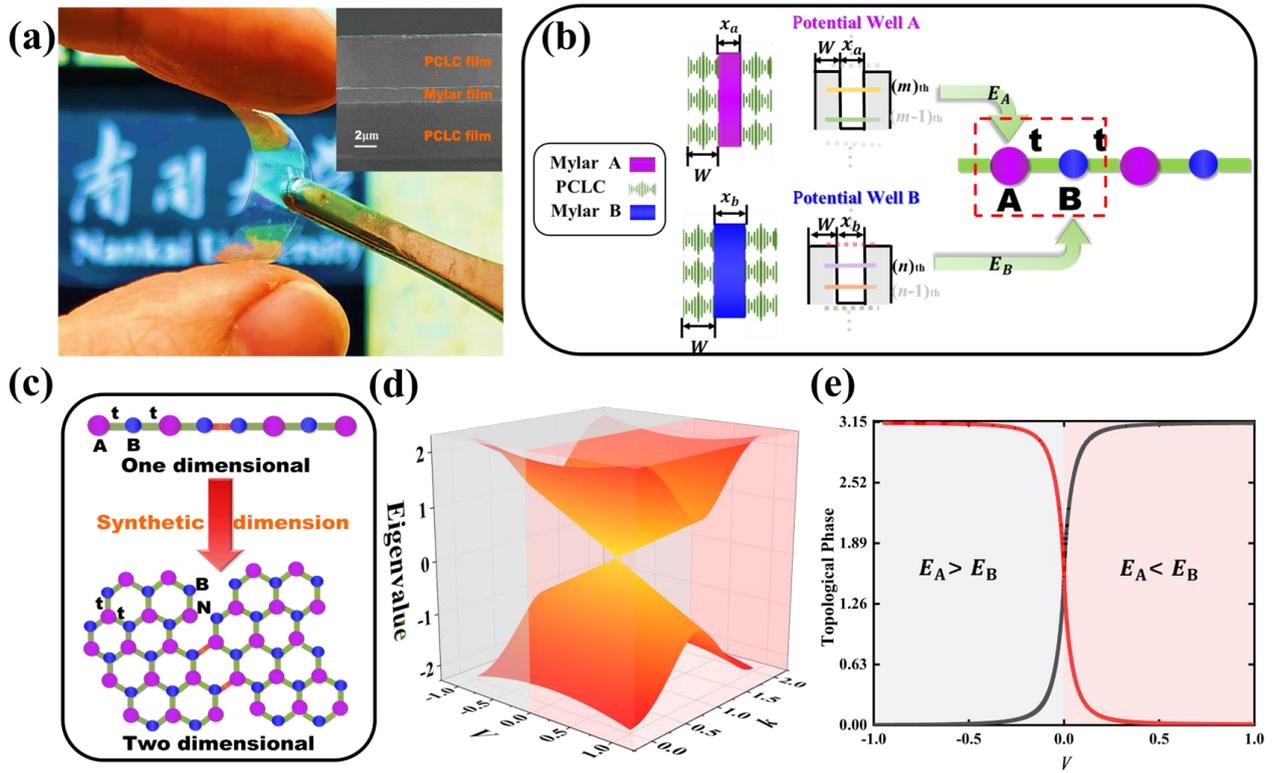

**Fig. 1. Illustration of basic principle to construct an inversion-symmetry broken flexible optical superlattice based on Semenov insulators.** (a) Photo of the free-standing three-layer structure consisting of an isotropic polymer layer (transparent film) sandwiched between two free-standing PCLC layers (transparent reflective films). The upper right inset shows the cross-sectional scanning electron microscope (SEM) image of this free-standing structure. (b) Schematic diagram of a 1D equally-coupled and on-site potentials modulated diatomic chain assembled by two kinds of optical potential wells (Mylar films). (c) Schematic diagram of synthetic dimension construction. By introducing the on-site potential modulation $V$, this diatomic chain can be equivalent to a 2D Semenov insulator, such as the BN crystal. (d) Dispersion relation in the $V$-$k$ synthetic parametric space. (e) Plot of phase changes as a function of $V$ in the range [-1, 1]. Red and black lines represent winding numbers of two bands, respectively.

Based on the optical quantum-well theory, we first analyze the 1D optical superlattice composed of an isotropic polymer layer (transparent film) inserted between two free-standing PCLC layers (transparent reflective films), as shown in Fig. 1a. The multilayer film was cut and sprayed with gold powder so as to observe its cross-sectional SEM image, as illustrated in the up-right inset. Studies have shown that one or more defect modes can be introduced into the polarization band gap of the CLC by adding isotropic layers and/or pitch jumps within the CLC helix [75]. Here, the PCLC has the characteristics as 1D photonic crystals, and its polarization band gap limits the propagation of photons with specific frequencies and polarizations, so it can be regarded as a potential barrier. A cheap, commercial and isotropic Mylar film (Mylar D, DuPont), inserted into the PCLC, can be regarded as a potential well to form an optical quantum well [76, 77]. The width of the single potential well is so narrow that the motions of the photons along the direction perpendicular to the well walls appear quantized characteristics and the energies take discrete values in the polarization band gap of the PCLC. Photons with different energies occupy discrete energy levels belonging to different orders, forming a series of photon bound states or quantum well states in the form of standing waves. Similar to the semiconductor quantum wells, the frequencies of the photon bound states will become lower with the increase of the well width [78, 79, 80]. [see Note 1 of Supporting Information (SI)].

For the diatomic chain shown in Fig. 1b, different thicknesses of Mylar films cause the change of the on-site potentials, which in turn change the coupling energy levels in adjacent potential wells, breaking the inversion symmetry of the system. Here, we consider only the nearest neighbor coupling, that is, one set of energy levels that can be coupled in the adjacent A and B quantum wells, such as the $m^{\text{th}}$ energy level in the A potential well and the $n^{\text{th}}$ energy level in the B potential well. The corresponding energies are defined as $E_A = -V$ and $E_B = V$, which can also be considered as the on-site potentials of A and B wells according to Fig. 1b. Here, $M = E_A - E_B$ describes the on-site potential energy difference between the wells. Then we can construct a 2D synthetic parametric space by means of the wave vector $k$ and the modulated on-site potential $V$. This 1D optical superlattice based on the equally-coupled and site-potential modulated diatomic chain can be likened to a 2D Semenov insulator, as shown in Fig. 1c.

The chain consists of $N$ unit cells, and each unit cell hosts two sites (potential wells A and B). According to the tight binding approximation, the tunneling coupling term between the nearest neighbor potential wells can be written as:

$$c = -\langle \phi_A(x)|H|\phi_B(x)\rangle \tag{1}$$

where $\phi_A(x)$ and $\phi_B(x)$ represent the eigenmodes of potential wells A and B, respectively. The Hamiltonian in real space can be expressed as

$$H = \sum_j c\big(a_j^\dagger b_j + b_j^\dagger a_{j+1} + h.c.\big) + E_A a_j^\dagger a_j + E_B b_j^\dagger b_j \tag{2}$$

where $a_j$ ($a_j^\dagger$) and $b_j$ ($b_j^\dagger$) are the annihilation (creation) operators in potential wells A and B in the $j^{\text{th}}$ unit cell of the lattice. By solving the eigenvalues corresponding to the Hamiltonian represented by equation (2), new energy levels formed by the resonant tunneling effect can be obtained. According to the Einstein-de Broglie relation [81], each energy level after splitting can be converted into the corresponding frequency, which can be verified by the transmission spectrum. Taking the synthetic parametric space into account, the Hamiltonian of this system in momentum space can be expressed more generally as:

$$H(V,k) = \begin{bmatrix} -V & c + ce^{-ik} \\ c + ce^{ik} & V \end{bmatrix} \quad (3)$$

Then we can get the energy eigenvalues corresponding to this Hamiltonian in the $V - k$ synthetic parametric space. Similar to the discussion in the 2D BN and other Semenov insulators where the potential difference is applied to the diagonal elements of the Hamiltonian matrix, we pay more attention to the effect of the on-site potential difference $M$ on the Hamiltonian of the system. If we only consider the coupling of the energy levels belonging to the same order in the adjacent potential wells, it can be found that a degenerate Dirac point appears if and only if $V = 0$, that is, the potential difference $M$ of A and B is zero, as shown in Fig. 1d. When $V \neq 0$, that is, $M > 0$ or $M < 0$, there is a band gap without any topological edge state inside the gap, which is consistent with Semenov insulators. It should be noted that the transition from $M > 0$ to $M < 0$ cannot be continuous. The system must go through the closure of the band gap and may lead to the appearance of the nontrivial interface state.

However, our optical superlattice is a multi-energy level system. Energy levels with the same energy may correspond to different orders, as shown in Table S2 of the SI. It can be found that there are a group of Mylar films with different thicknesses corresponding to the same wavelength of 574 nm, that is, potential wells with different widths ($M \neq 0$) can have different energy levels corresponding to the same energy. The quantum size effect and multi-level characteristics of quantum wells ensure the condition for band gap closure to be $M = 0$, as shown in the Note 2 of the SI.

In order to find the topological invariant of this 1D superlattice, the Hamiltonian is rewritten in the form of Pauli matrix as $H = d_x\sigma_x + d_y\sigma_y + d_z\sigma_z$. The coefficients of the Pauli matrix can be expressed as:

$$d_x = c + c\cos(k), \quad d_y = c\sin(k), \quad d_z = V \quad (4)$$

Despite the simple form of the Hamiltonian, it has described a number of physical systems in condensed-matter physics for which topological phase effects have been discussed [82, 83, 84]. The eigenvalues of $H$ are given by $E_\pm = \pm|\vec{d}|$, where $\pm$ corresponding to two non-degenerate bands caused by the on-site potential difference. The eigenstates are unit vectors along or against the $\vec{d}(k)$ direction, and the winding number can be quantitatively expressed as [85, 86, 87]:

$$\gamma^\pm = \frac{1}{2\pi} \int_0^{2\pi} \frac{1}{E_\pm(E_\pm - d_z)} (d_x \vartheta_k d_y - d_y \vartheta_k d_x)\, \mathrm{d}k \tag{5}$$

According to equation (5), the winding number of a single band changes as $V$ varies in the range $[-1,1]$ as shown in Fig. 1e, where the red and black lines represent $\gamma^+$ and $\gamma^-$ respectively. The gray ($V < 0$) and red ($V > 0$) areas denote the reverse of on-site potentials of A and B. Thus, the sub-structures from the gray and red areas shown in Fig. 1d, e have different topological properties.

**Topological interface states in soft-matter based topological VCSEL**

Here, we choose 3 μm- thick Mylar films A and 4 μm-thick Mylar films B to construct two sub-structures with different topological properties (Fig. 2a). The on-site potentials in the sub-structures of gray and red areas satisfy $E_A > E_B$ and $E_A < E_B$, respectively. The topological phase transition process can also be verified according to the symmetry of the electric field distributions of the eigenstates above and below the common band gap at the edge of the first Brillouin region, as shown in the Note3 of the SI. Here, potential wells A and B have similar energies near the wavelength of 575 nm, while belong to the $m^{\text{th}}$ and $n^{\text{th}}$ energy levels respectively. Due to the tunneling and coupling between potential wells A and B, the original degenerate energy levels split, and a series of splitting resonance peaks appear in the transmission spectrum [78, 79, 80]. In turn, mini-bands are formed and mini-band gaps are generated. This phenomenon can also be analyzed according to the tight binding method [88].

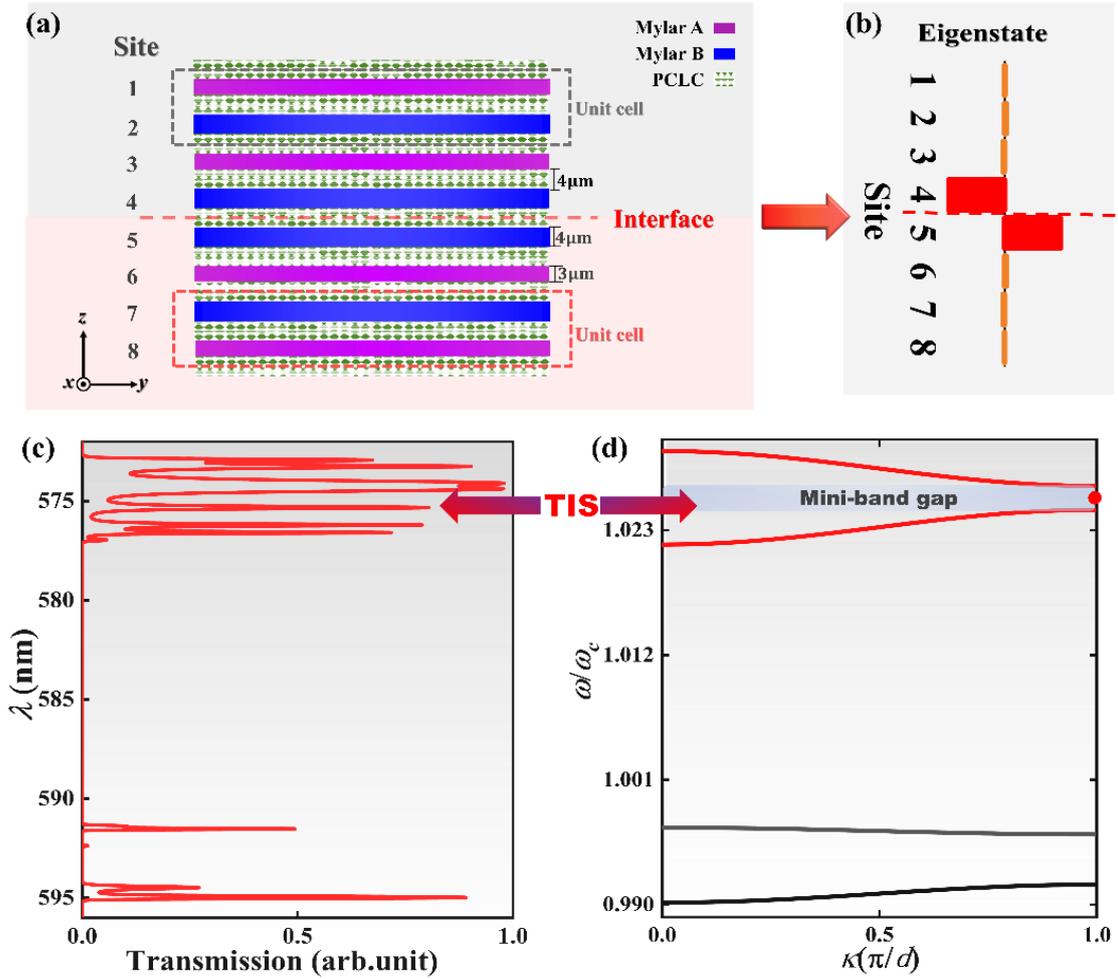

**Fig. 2. TIS in an optical superlattice used for realizing soft-matter-based topological VCSELs.** (a) 1D topological VCSEL composed of two sub-superlattices with different topological properties. (b) Eigenstate distribution of the TIS. (c) Transmission spectrum and (d) mini-band structure of the topological optical superlattice.

It has been proven in BN and SiC crystals that splicing two structures chosen before and after the Dirac point is closed, namely symmetrically selecting two structures from the gray and red areas, can generate a robust topological interface state (TIS) [68, 69]. Thus, by splicing two topologically unequal sub-structures, a TIS which is highly localized in the isotropic layers on both sides of the splicing interface will be generated, as shown in Fig. 2b. From the transmission spectrum and the mini-band structure of the spliced superlattice, it can be found that the TIS originates from one of the bulk states and evolves into the mini-band gap, that is, one bulk state turns into the TIS, as indicated by the arrows in Fig. 2c, d. In addition, because the energy levels $E_{m-1}$ and $E_{n-1}$ inside the two potential wells A and B have a large energy difference, the photons between the nearest potential wells cannot tunnel and couple. As a result, we cannot find any TIS within 590 nm - 598 nm.

From the perspective of application, the most valuable advantage of the TIS should be its robustness against disorder and defect, which has been discussed in detail based on the SSH model with chiral symmetry protection [89]. In addition, some studies have demonstrated that localization of the topological state can be maintained even if the chiral symmetry is broken by directly perturbing the on-site potentials [90]. From the calculation results in Note4 of the SI, we can find that the TIS basically keeps its corresponding eigen-wavelength unchanged until the disturbance factor $\gamma$ is equal to about 0.25. At the same time, the strong localization of the TIS is not affected by the degree of disorder. However, the bulk states change almost from the beginning of the disordered disturbance. Therefore, compared with bulk states, the TIS is more robust against the disorder disturbance on the on-site potentials.

**Lasing characteristics of soft-matter-based topological VCSEL**

Here, we prepared the soft-matter-based optical superlattice by vertically stacking left-handed PCLC films and commercial Mylar films, without any complex lithography or deposition technique. The stacking diagram and the cross-sectional SEM image of one unit cell are shown in Fig. 3a, b. It consists of two Mylar films of different thicknesses and three PCLC films of the same thickness. As can be seen from the Fig. 3b, layers are stacked flat and closely. By spin-coating a gain medium PM597 of negligible thickness on each PCLC film, a flexible, unsupported, circularly polarized topological VCSEL can be achieved. Fig. 3c1, c2 shows the photographs of two superlattices with and without PM597 respectively, consisting of 17 layers on a flexible PET substrate according to the superlattice structure shown in Fig. 2a. The areas framed by the red dashed boxes are the effective regions. Since each film itself was not very sticky, in order to prevent loosening of the sample during the experiment, we used a transparent tape to seal the edges around the samples.

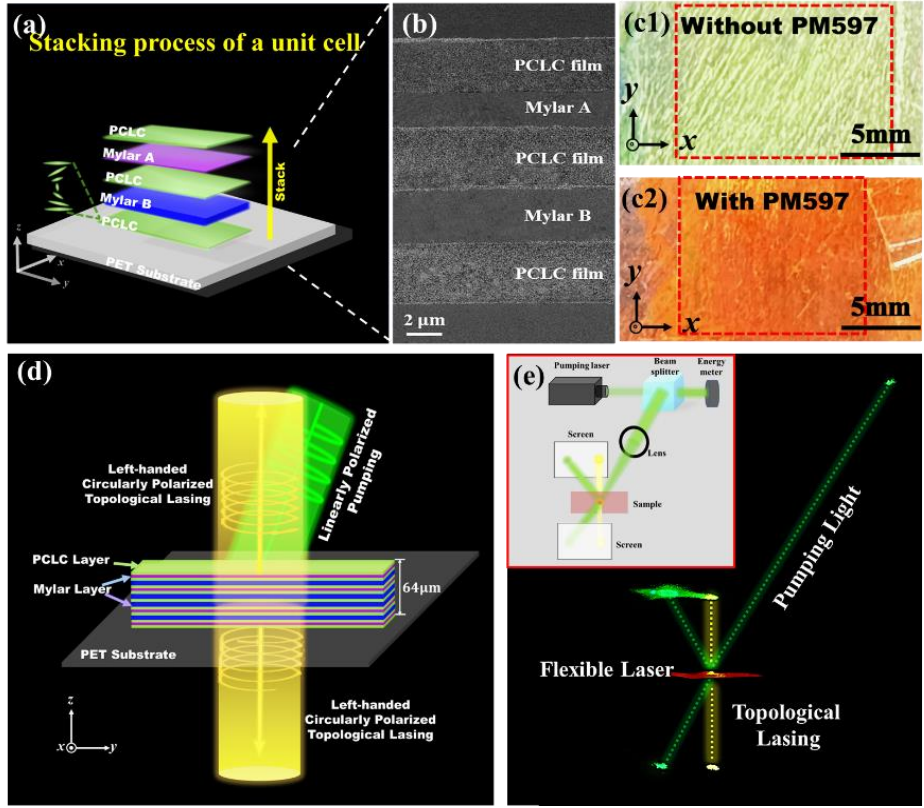

**Fig. 3**. **Laser emission from the soft-matter-based topological VCSEL.** (a) Schematic diagram of stacking one basic unit cell with five layers. (b) Cross-sectional SEM image of one basic unit cell of the topological VCSEL. (c) Pictures of the optical superlattices consisting of 17 layers without (c1) and with (c2) the dye PM597. The areas marked by the red dashed frames are the effective regions of the samples. (d) Diagram of the dual-side lasing when pumped by linearly polarized light. The thin film is parallel to the *x-y* plane, and the lasing emissions are along the *z* and -*z* directions. (e) A photograph of actual lasing patterns output from two sides of the sample. Due to the fluorescent dye PM597, the film appears purplish red. Yellow light spots at the direction perpendicular to the sample represent the left-handed circularly polarized TIS lasing, while green light spots represent the linearly polarized pump light. The inset shows the corresponding experimental setup.

Then, we used a second harmonic output of a Q-switched Nd: YAG laser (SLIII-10, Continuum) with a wavelength of 532 nm, a repetition rate at 10.0 Hz, and a pulse duration of 4.0 ns to excite the sample. Then, left-handed circularly polarized topological lasing at 575 nm could be generated from this soft-matter-based topological VCSEL, as shown in Fig. 3d. Two yellow laser beams pointed in opposite directions perpendicular to the surface. The specific experimental setup is shown in the inset in Fig. 3e. In order to clearly show the paths of the pump beam and the topological lasing beam, we use dotted lines to connect the lens to the sample, and the sample to the spots on the light screens in Fig. 3e.

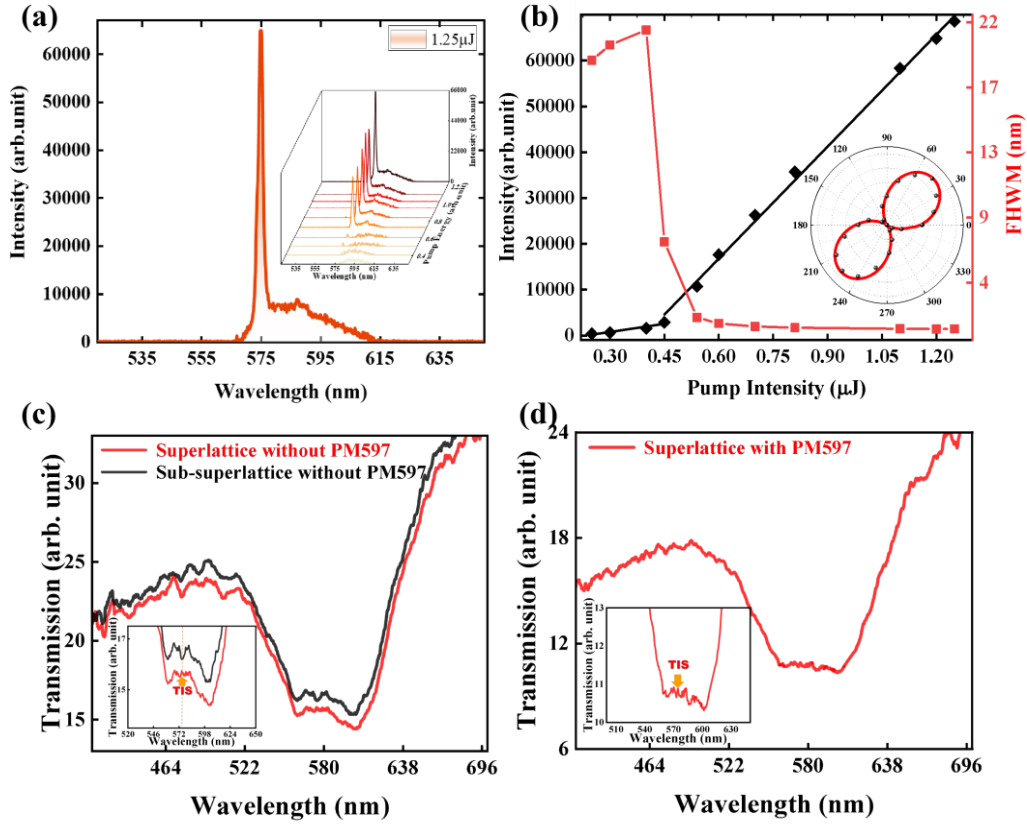

**Fig. 4. Measured emission and transmission spectra of the topological VCSEL.** (a) Emission spectrum obtained from the superlattice pumped at 1.25 μJ. The inset shows the emission spectra at different pump energies. (b) Plot of the peak intensity and linewidth of the TIS emission as a function of the pump energy. The inset shows the intensity of the quarter-waveplate-transformed laser radiation as a function of the polarization angle in polar coordinates. The polar angle stands for the transmission angle of the polarizer, and the radius stands for the transmittance. (c) Transmission spectra of the sub-superlattice (black) and spliced superlattice (red) without dye PM597. (d) Transmission spectra of the spliced superlattice with dye PM597. The insets are zoom-in spectra near the TIS lasing wavelength.

We further examined the topological lasing threshold and polarization characteristics through the experimental setup shown in Materials and Methods. Fig. 4a shows the single-mode TIS lasing with a peak wavelength of 575 nm at a pump energy of 1.25 μJ, and the inset shows the emission spectra under different pump energies. From Fig. 4b, it can be found that when the pumping energy exceeds 0.47 μJ ($1.5 \times 10^6$ W/cm$^2$), the full width at half maximum of the emission peak is obviously narrowed, indicating the lasing behavior. This threshold is significantly lower than those of the current topological VCSELs [47, 48, 49, 57, 58]. In addition, as the pump energy is larger than 4.4 μJ, other defect states in the superlattice will also be excited. Our experimental results show that the threshold of the TIS lasing is significantly lower than that of other defect states because the TIS in the superlattice has the highest photonic density of state. A quartz quarter-wave plate with its slow axis parallel to the vertical axis and a polarizer were used to examine the polarization characteristics of the TIS lasing. As shown in the inset of Fig. 4b, when the polarizer is rotated to 45° and 225° relative to the horizontal axis, the transmitted intensity is maximum, while the lasing is completely wiped out at 135° and 315°, indicating that the TIS lasing emitted from the topological VCSEL has a good left-handed circular polarization characteristic. Besides, its slope efficiency reaches 4.18%, and its total lasing conversion efficiency (the ratio of lasing pulse energy to pump pulse energy) reaches 3.2% under the pump energy of 2.0 μJ, which is much higher than that of PCLC lasers [91].

In addition, we also measured the transmission characteristics of the sub-superlattice without the splicing interface and the spliced superlattice in Fig. 2a with and without dye PM597, as shown in Fig. 4c, d. Here, a deuterium halide natural white light source (DH-2000-BAL, Ocean Optics) was used to irradiate the sample through a lens with a focal length of 85 mm for measuring the transmission spectrum. The transmitted light, after passing through the sample, was collected by another lens with the same focal length and coupled to a spectrometer (HR4000CG-UV-NIR, Ocean Optics). However, due to the self-selective reflection and unavoidable light scattering of the PCLC films, the transmittance of the stacked multilayer structure will be greatly reduced. Therefore, the multilayer structures formed by PCLC films in our current experimental studies have a limited number of layers. Also limited by the resolution of the spectrometer, we could not observe sharp transmission peaks for the time being, but we could still find several transmission peaks corresponding to the TIS with wavelength of 575 nm and other bulk resonant modes in the transmission spectra of the spliced superlattices with and without dye PM597. From Fig. 4c, we can clearly see that TIS is located in the center of the mini band gap. These results are in good agreement with our theoretical calculations, as shown in Fig. 2c. Since the band gap of the PCLC we used ranges from 560 nm to 604 nm, the samples reflect yellow light.

**Two application examples of flexible topological VCSEL**

Interestingly, our topological VCSEL can not only maintain good single longitudinal mode emission under low pump intensity, but also carry the special transverse mode distribution of the pump light, i.e., its transverse mode is decided by the pump light. A photomask engraved with two letters 'NK' was inserted into the pump light path and imaged on the sample by a lens with a focal length of 85 mm. The laser intensity distribution was monitored by a laser beam profiler (SP620U, Ophir-Spiricon) through an objective lens, as shown in Fig. 5a. As illustrated in Fig. 5b, the pump beam reflected by the sample and the lasing beam can both be displayed on a screen. Fig. 5c2 clearly shows the transverse profile of the topological laser presented on the laser beam profiler, which corresponds well to the photomask shown in Fig. 5c1. Therefore, our topological VCSELs have great potential to be used in optical display devices.

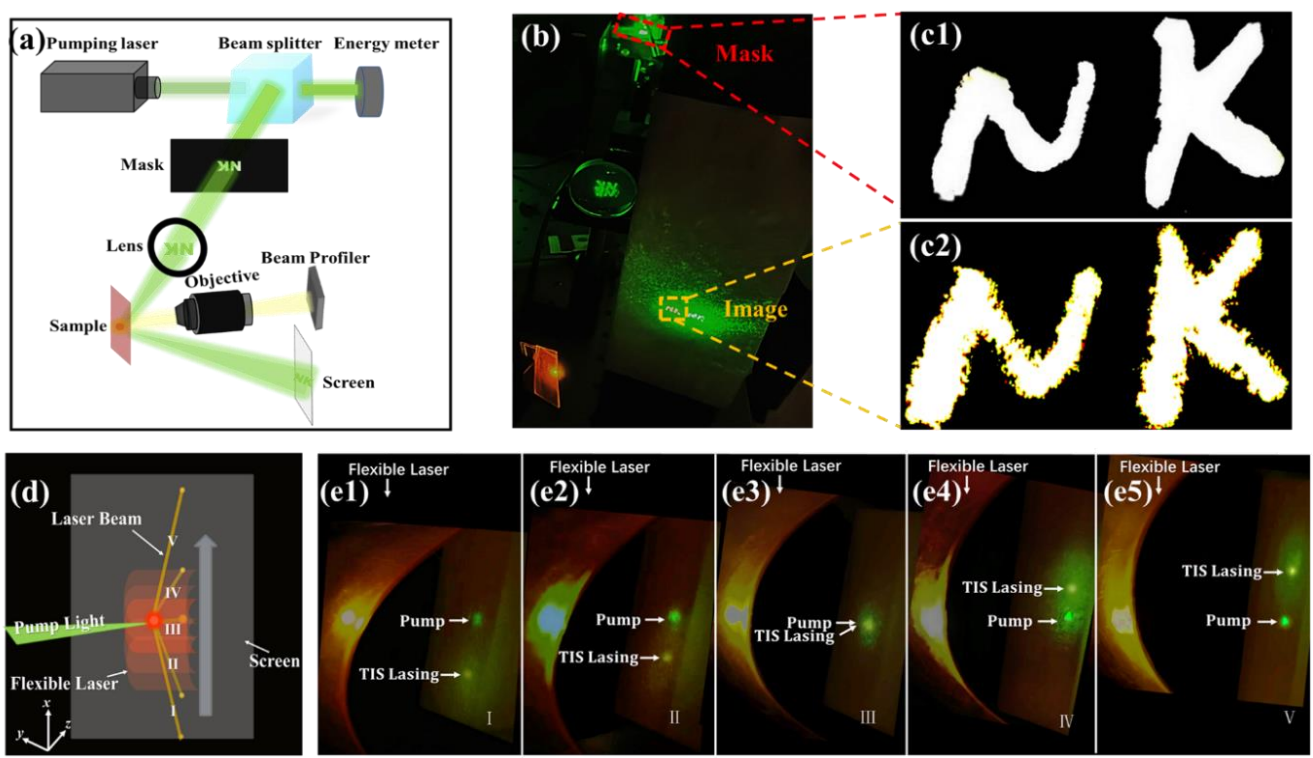

**Fig. 5**. **Demonstration of two application examples with this flexible topological VCSEL.** (a) Experimental setup and (b) photograph demonstrating a topological VCSEL excited by a pump beam passing through a photomask containing the letters 'NK'. (c) Photographs of the photomask (c1) and the transverse profile collected by the beam profiler (c2). The black area of the mask is opaque, while the white area of 'NK' is transparent. (d) Diagram of the laser beam steering caused by moving the curved sample. (e1-e5) Experimental photographs taken when the curved sample is moved from bottom to top, and thus the laser illuminates at different positions of the sample (I-V).

To demonstrate the flexibility of the topological VCSEL, we designed an experiment in which the laser beam could be diverted by moving a curved topological VCSEL film, as shown in Fig. 5d. In this experiment, we bent the sample into a cylindrical surface by fixing it on a cylinder, and moved the cylindrical sample from bottom to top. As a result, the pump beam of the Nd:YAG laser would illuminate different regions of the cylindrical film, so that five different locations on the film were pumped, labeled as I-V. The deformation of the film caused the helical axes of the PCLCs along different normal directions of the cylindrical surface, which in turn caused the outgoing lasing spots to fall on the screen at different locations. As shown in Fig. 5e1-e5, the corresponding TIS lasing spots (yellow spots) and pump lasing spots (green spots) could be seen on the screen, and these images showed the beam steering phenomenon of the flexible topological VCSEL. Adjusting the radius of curvature of the VCSEL could control the beam steering degree. This property means that the topological VCSEL can lase in a range of directions without rotating the laser device. Notably, the topological VCSEL is thermally stable and retains its original lasing characteristics after fourteen months of storage and thousands of bends, as shown in the Note5 of the SI. These characteristics of such laser devices are particularly attractive for potential wearable photonic technologies and compatible platforms.

**Discussion**

In general, we have designed and demonstrated soft-matter-based optical superlattices consisting of PCLC films with fixed thickness and Mylar films with two different thicknesses, and achieved a circularly polarized flexible topological VCSEL in the visible wavelength region for the first time. Due to the quantum size effect of optical quantum wells, we can adjust the spacing and number of energy levels by changing the width of the well, and then change the on-site potentials in the superlattice. Based on the synthetic parametric space constructed by the on-site potential modulation and the original geometric parameters, the topological phase transition process in our inversion symmetry broken 1D optical superlattice can be analyzed, which is similar to 2D BN lattices or other Semenov insulators. It is proved that the TIS located at the interface can still be generated by splicing two inversion symmetry broken 1D sub-superlattices with different topological invariants. After introducing the gain medium into the superlattice, a single longitudinal mode TIS lasing with a wavelength of 575 nm and a threshold of 0.47 μJ ($1.5 \times 10^6$ W/cm$^2$) can be observed under pulsed laser pumping, which can also keep the transverse profile of the pumped light. In particular, this topological VCSEL also has a higher laser conversion efficiency and slope efficiency compared to the existing PCLC lasers. Lasing beam steering is demonstrated without tilting the laser device. This soft-matter-based topological VCSEL has extremely low production costs, does not require complex processing technology, and is easy to integrate on any substrate. Thus, it is of great significance to promote the practical applications of topological photonic devices in the fields of wearable devices, artificial intelligence, information encryption, consumer electronics and other fields.

**Materials and Methods**

**Fabrication of the soft-matter-based topological VCSEL.**

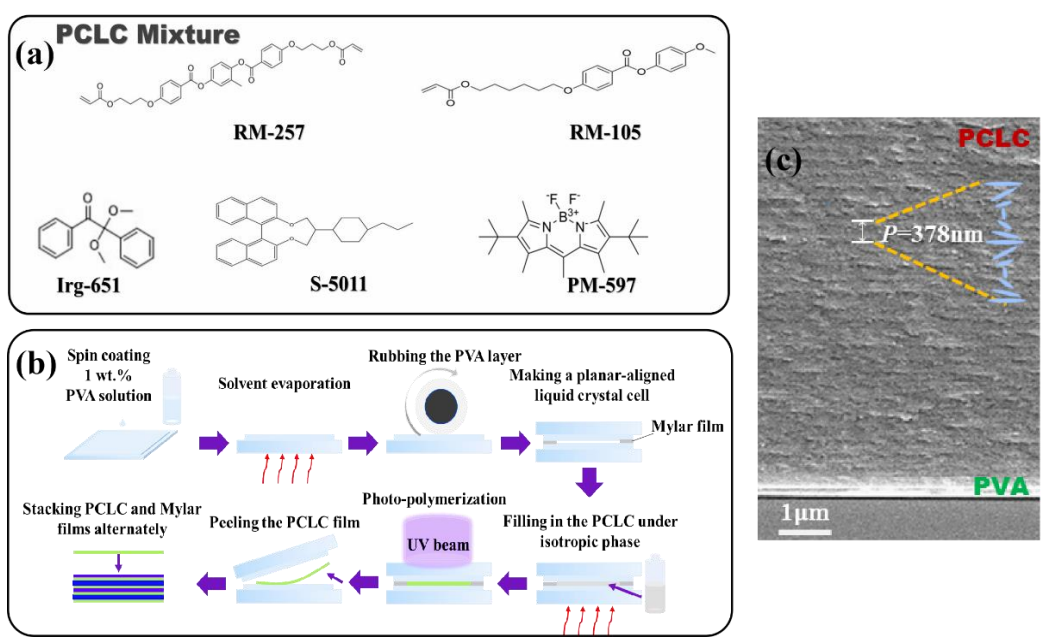

**Fig. 6**. **Preparation process of the soft-matter-based topological VCSEL.** (a) Structures of molecules used for making PCLC films. (b) Schematic diagram of the fabrication process of the topological VCSEL. (c) Cross-sectional SEM image of the PCLC film. The PCLC has a periodic structure along $z$ direction with a pitch of $P$.

In order to achieve laser emission from the vertical cavity surface, the helical axis of the PCLC was designed to be perpendicular to the substrate, i.e., along the $z$ direction. Thus, we prepared anti-parallel plane-oriented liquid crystal cells by friction orientation technology, and the orientation of liquid crystal molecules was realized by rubbed polyvinyl alcohol (PVA, Sigma-Aldrich) alignment layers. Here, the polymerizable liquid crystal mixture contained 73 wt.% RM257 (JCOPTIX) and 24.5 wt.% RM105 (JCOPTIX), which together formed a basic nematic mixture. Then, 2.05 wt.% left-handed chiral dopant S5011 with a high torque coefficient and 0.45 wt.% photo initiator

Irgacure-651 (BASF) were added as shown in Fig. 6a. These mixtures were filled into home-made planar LC cells. The thickness of the LC cells was controlled by a 4 μm-thick Mylar film as the spacer. In order to make the PCLC form a uniform helical alignment, the sample was cooled from the isotropic phase slowly, and a clear photonic band gap of the PCLC from 560 nm to 604 nm could be seen from its transmission spectrum. A UV curable source (HTLD-4Height-LED Opto-electronic Tech Co., Ltd) with a wavelength of 365 nm and an intensity of 100 mW/cm² was used to irradiate the sample from a distance of 20 cm. In order to ensure uniform curing, both sides of each LC cell were irradiated for 5 minutes. Then, the glass substrates of the LC cell were torn apart, and the independent PCLC film could be peeled off with a sharp tweezer, as shown in Fig. 6b. Fig. 6c is a cross-sectional SEM image of one detached PCLC film. Here, PCLC shows a clearly periodic structure along the $z$ direction. Then, a flexible topological VCSEL could be constructed by alternately stacking the detached PCLC films with commercial Mylar films.

**Experimental setup for lasing measurements**

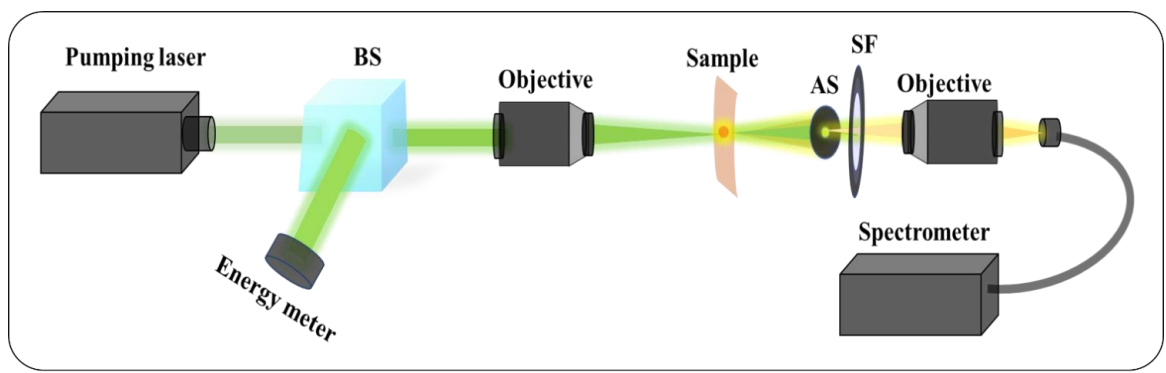

**Fig. 7. Illustration of the experimental setup for investigating the lasing characteristics of topological VCSELs.** BS: beam splitter, AS: aperture stop, SF: spectral filter.

Here, we examined the topological lasing threshold and polarization characteristics through the experimental setup shown in Fig. 7. The pump light was divided into two beams by a beam splitter. An energy meter (LabMax-Top, Coherent Inc.) was used to monitor the pumping energy in the reflection direction. An objective lens with a focal length of 85 mm was used to focus the beam in the transmitted direction onto the sample. To avoid the influence of the pump laser and protect the spectrometer, a 532 nm notch filter was placed behind the sample. The TIS lasing was collected by a 10× objective lens. A pinhole with a diameter of 1 mm was placed between the sample and the filter to block stray light. The light passing through another objective was coupled to an optical fiber of a high-resolution spectrometer (SP2358, Princeton Instruments).

**Acknowledgments**

This work is supported by the National Key Research and Development Program of China (2022YFA1404800), National Natural Science Foundation of China (12074201, 12134006), Key project of Tianjin Natural Science Foundation (23JCZDJC00920), 111 Project (B07013), PCSIRT (IRT_13R29), the Fundamental Research Funds for the Central Universities, Nankai University (63241602), Postdoctoral Fellowship Program of CPSF under Grant Number 20240758 and Slovenian Research Agency research program P1-0192.